\documentclass[12pt]{article}
\usepackage{epsfig}
\usepackage{amsfonts}
\usepackage{amscd}
\usepackage{latexsym}
\usepackage{amsmath,amssymb}
\usepackage{verbatim}
\usepackage{setspace}
\usepackage{color}
\usepackage{cite}

\usepackage[textheight=9in, textwidth=6.5in, letterpaper]{geometry}

\usepackage{color}   %May be necessary if you want to color links
\usepackage{hyperref}
\hypersetup{
    colorlinks=true,  %set true if you want colored links
    linktoc=all,     %set to all if you want both sections and subsections linked
    linkcolor=black,  %choose some color if you want links to stand out
    citecolor=black,
    filecolor=black,
    urlcolor=black,
}

\numberwithin{equation}{section}

\def\p{\partial}
\def\cl{{\cal L}}

\def\<{\langle}
\def\>{\rangle}

\def\cO{\mathcal{O}}

\def\t{\tau}

\def\be{\begin{equation}}
\def\ee{\end{equation}}
\def\beq{\be\begin{array}{c}}
\def\eeq{\end{array}\ee}
\def\bes{\be\begin{split}}
\def\ees{\end{split} \ee}
\def\bs{\begin{split}}
\def\es{\end{split} }

   \makeatletter
  \let\over=\@@over \let\overwithdelims=\@@overwithdelims
  \let\atop=\@@atop \let\atopwithdelims=\@@atopwithdelims
  \let\above=\@@above \let\abovewithdelims=\@@abovewithdelims
\renewcommand\section{\@startsection {section}{1}{\z@}%
                                   {-3.5ex \@plus -1ex \@minus -.2ex}%nn
                                   {2.3ex \@plus.2ex}%
                                   {\normalfont\large\bfseries}}

\renewcommand\subsection{\@startsection{subsection}{2}{\z@}%
                                     {-3.25ex\@plus -1ex \@minus -.2ex}%
                                     {1.5ex \@plus .2ex}%
                                     {\normalfont\bfseries}}

\begin{document}
\begin{titlepage}
\unitlength = 1mm

\vskip 1cm
\begin{center}

{ \LARGE {\textsc{Dual Fluid for the Kerr Black Hole}}}

\vspace{0.8cm}
 Vyacheslav Lysov

\vspace{1cm}

{\it Walter Burke Institute for Theoretical Physics,\\
 California Institute of Technology, \\
 Pasadena, CA 91125, USA}

\begin{abstract}
Rotating black holes are algebraically special solutions to the vacuum Einstein equation. Using properties of the algebraically special solutions we construct the dual fluid, which flows    on black hole horizon. 
An explicit form of the Kerr solution allows us to write an explicit dual fluid solution and investigate its stability using energy balance equation.  We show that the dual fluid is stable 
because of high algebraic speciality of the Kerr solution. 

\end{abstract}%\vspace{0.5cm}

\vspace{1.0cm}

\end{center}

\end{titlepage}

\pagestyle{empty}
\pagestyle{plain}

\pagenumbering{arabic}

\tableofcontents

\section{Introduction}

The fluid/gravity correspondence is a framework that connects the knowledge about the gravity equations and fluid dynamics. There are multiple approaches to such correspondence  originating from membrane paradigm \cite{Price:1986yy} , AdS/CFT \cite{Bhattacharyya:2008jc,Bhattacharyya:2008kq}, shear perturbation\cite{bkls,Bredberg:2011jq}, quasinormal modes\cite{Son:2007vk,Baier:2007ix}, algebraic speciality \cite{Lysov:2011xx}   and others.
Each approach describes an explicit pair for  the gravity and fluid equations with an additional mapping procedure for the solutions.  In our paper we want to make a step further and use the fluid/gravity correspondence to answer interesting questions about the dual fluid in terms of the   geometric data. The two major questions we are attempting to approach are the {\it explicit solutions} and {\it solutions stability}.  Both questions are interesting and natural for the  nonlinear equations 
on both fluid and gravity sides. An ability to write a solution in closed form often teaches us about some new features of the solution such as symmetries  or algebraic speciality.  Thus an idea to apply the fluid/gravity mapping to 
know closed form solution on either side to generate more closed form solutions  is very attractive.  Another advantage of an explicit   solutions is an ability to investigate the stability of the solutions. In context of fluid dynamics the stability is closely related to the phenomenon of turbulence.

Among the multiple fluid/gravity mappings the algebraic speciality approach  fits best to answer questions that we described. In our paper  \cite{Lysov:2011xx} we described the simplest setup: minimally special geometry and (conformally) flat hypersurface geometry. Unfortunately almost all  explicit solutions to Einstein equations   are more special \cite{Coley:2007tp}, i.e. have additional vanishing Weyl tensor components. Furthermore, the horizon geometry 
is far from being (conformally) flat for most solutions. Thus we need to modify our approach to adopt for the   known black hole solutions. While it is possible to describe the most general algebraic constrains and hypersurface geometry we 
decided to restrict our discussion to  the rotating black holes. Such restriction allows us to make discussion much simpler while    keeping several interesting phenomenon and being physically relevant on both gravity and fluid sides. 

The generic rotating black hole  geometry turns out to be very complicated to analyze straight on, so we start our discussion with slowly rotation case, where the angular momentum $J$ is much smaller then the mass squared $M^2$. 
The geometry is an algebraic  type {\bf D}, while the  $r=const$  fluid hypersurface preserves the smaller type {\bf I}$_i$ subset  of type {\bf D} constraints. The type    {\bf I}$_i$ constraints have type {\bf I} as a subset so we can 
construct the dual fluid, while the additional constraints restrict possible fluid solutions to Killing flows. Interestingly, the simple analysis of additional speciality constraints provides a quick way solve the Navier-Stokes (NS) equations and  to reproduce the slowly rotating  black hole results  from \cite{Bhattacharyya:2007vs}. Since slowly rotating Kerr solution  can be viewed as a small perturbation over the Schwarzschild   geometry our NS system and fluid solution match with the results by Bredberg and Strominger \cite{Bredberg:2011xw}. 

The fluid dual to the generic rotating black hole solution  obeys the modified version of an incompressible NS equation. The additional terms are similar to ones described in fluid/gravity generalizations \cite{Wu:2015pzg}. Similarly to the toy case 
higher algebraic speciality provides additional constraints on velocity  similar to the Killing equation. A solution to additional constraints provides a quick way of solving the NS equation for a dual fluid. As an example we describe in details both fluid equations and solution for the case of 4d Kerr geometry. 

Given an explicit form of the fluid equations and solutions we can investigate the linearized stability of the flow.  The most common approach to stability is based on the {\it Reynolds number} estimation. The Reynolds number for the slow rotating case is equal to the $3J/M^2$ and is parametrically small. We also estimate the critical Reynolds number, where fluid can be unstable, using the energy balance equation. It turns out that the 
dual fluid to the generic type {\bf{D}} black hole solution   is always stable.

We begin our discussion with the review of the algebraically special fluid/gravity correspondence in section 2.  In section 3 we review the Kerr geometry and describe the dual fluid hypersurface. In section 4 we    consider a slow rotating Kerr solution and apply the fluid/gravity map to it. In section 5 we describe the dual fluid to the Kerr black hole and discuss its stability in section 6. In the closing section  7 we discuss the results and possible generalizations.

\section{The fluid/gravity correspondence}

In this section we will review the fluid/gravity correspondence and modify it to include the rotating black hole solutions.   

Given a solution   to the Einstein equations  in $p+2$ dimensions and a  time-like codimension-one hypersurface $\Sigma$ we can construct a symmetric conserved two-tensor $T_{ab}$
\be\label{cov_cons}
\nabla^a T_{ab}=0,\;\;\; T_{ab} =T_{ba}, \;\;\; a,b =0,...,p, 
\ee 
where $\nabla^a$ is a covariant derivative with respect to the induced metric $h_{ab}$ on a hypersurface. The tensor $T_{ab}$ is often called the Brown-York (BY) tensor  and is constructed from the extrinsic curvature 
\be
T_{ab} \equiv K_{ab}-\frac12 h_{ab} K,\;\;\; K_{ab} = \frac12 \cl_n h_{ab},
\ee  
with $n$ being the unit normal to $\Sigma$ and $K \equiv K_{ab}h^{ab}$. The fluid/gravity correspondence maps the BY tensor into the fluid stress tensor.  Fluid equations can be  formulated in the form of covariant conservation for the fluid stress tensor. However the $(p+1)(p+2)/2$ components of the stress tensor are expressed in terms of $p$-components of the fluid velocity and two scalars: density and pressure. In case of gravity-dual fluid the additional constraints for the BY tensor come from metric being algebraically special. 

The covariant conservation (\ref{cov_cons}) originates from the bulk Einstein equation $G_{\mu\nu}=0$ $\mu,\nu=0,...,p+1$ on  the hypersurface $\Sigma$ with coordinates $x^a,\;\; a=0,...,p$
\be\label{covariant_cons}
0=G_{\mu b } n^\mu|_{\Sigma}  = \nabla^a T_{ab}=\nabla^a K_{ab}-\p_b K
\ee 
and
\be\label{hamiltonian_cons}
0=2G_{\mu\nu}n^\mu n^\nu|_{\Sigma}  = K^2 - K_{ab}K^{ab} -R^{(p+1)}, 
\ee
where we use $(p+1)$ upperscript for the hypersurface quantities. The minimally special metrics are algebraic type {\bf I}, what imposes a constraint
\be\label{type_I_gen}
C_{{\bf l} i {\bf l} j}\equiv C_{\mu\nu\lambda\rho}\; {\bf l}^\mu\; m^{\nu}_i \; {\bf l}^\lambda\; m^\rho_j=0,\;\;\; i,j=1,...,p,
\ee  
where $C_{\mu\nu\lambda\rho}$ is the Weyl tensor and   ${\bf l},{\bf k}, m_i$ are the null frame vectors. The number of additional constraints in (\ref{type_I_gen}) is just enough to make equal the number of BY tensor components  and the total number of fluid equations (\ref{covariant_cons}),(\ref{hamiltonian_cons}).  

We often have a solution to Einstein equation and a null frame where solution is maximally special.  The rotational black holes are type {\bf D} solutions in four dimensions and type {\bf II}$_i$ in higher dimensions. 
Type {\bf II}$_i$  contains type {\bf I} as a subset, so we have freedom of choosing the  hypersurface $\Sigma$ and its time-foliation to preserve 
some subset of algebraic constraints.   In particular we want the null frame  vectors   to be 
\be\label{canonical_frame}
\sqrt{2}\;{\bf l} = T-n,\;\;\; \sqrt{2}\;{\bf k} = T+n,  
\ee 
with $T$ being unit normal for the time foliation $\Sigma_T$ of $\Sigma$. The rest of the null frame vectors $m_i$ are orthonormal basis for tangent space to $\Sigma_T$. 

We can express the  $p+2$-dimensional Riemann tensor on a hypersurface in terms of intrinsic geometry  and extrinsic curvature $K_{ab}$
\beq\label{riemann_on_surface}
R_{abcd} = R^{(p+1)}_{abcd}- K_{ac}K_{bd}+K_{bc}K_{ad},  \\ 
R_{anbc}= \nabla_b K_{ac} - \nabla_c K_{ab},   \\
R_{anbn}=  R_{ab} -R^{(p+1)}_{ab} +  KK_{ab} - K_{ac}K^c_{~b},  
\eeq
to evaluate 
\beq\label{type_I}
2R_{{\bf l}i{\bf l}j} = -R^{(p+1)}_{ij} +R_{ij}+  KK_{ij} - K_{ic}K^c_{~j}    \\ 
+R^{(p+1)}_{TiTj}- K_{TT}K_{ij}+K_{iT}K_{Tj} -  \nabla_i K_{jT} - \nabla_j K_{iT}+2 \nabla_T K_{ij}  =0
\eeq

The type {\bf I} constraint (\ref{type_I_gen}) is written for the Weyl tensor components so generally we need to subtract the trace from the Riemann tensor expression (\ref{type_I}). By construction our metrics solve Einstein equations so any traces of Riemann tensor vanish. However in order to avoid counting the same equations twice we can drop the  Hamiltonian constraint (\ref{hamiltonian_cons}) and use the type {\bf I} constraint in the form  (\ref{type_I}).  The equation (\ref{type_I}) can be used to solve for $K_{ij}$ in terms of $K_{iT}$ with solution becoming much simpler for the large mean curvature hypersurfaces. The natural candidates for the large mean curvature  hypersurfaces are black hole horizons.

\section{Kerr geometry} 
The Kerr metric in Boyer - Lindquist  (BL) coordinates is of the form 
\beq\label{Kerr_BL}
ds^2 = -\left(1-\frac{2Mr}{\rho^2}\right) dt^2 - \frac{4Mar\sin^2\theta}{\rho^2}dt d\phi +\frac{\rho^2}{\Delta} dr^2
+\rho^2 d\theta^2 +\left(r^2+a^2+\frac{2Ma^2r \sin^2\theta}{\rho^2}\right) \sin^2\theta d\phi^2 \\
\rho^2 = r^2+a^2\cos^2 \theta,\;\;\; \Delta = r^2+a^2-2Mr.   
\eeq
The Kerr metric is an algebraically special metric of type {\bf D}. The null frame in BL coordinates is given by the Kinnersley tetrad
\beq\label{type_D_frame}
{\bf k} =\frac{r^2+a^2}{\Delta}\p_t +\p_r +\frac{a}{\Delta}\p_\phi, \\
 {\bf l} =\frac{r^2+a^2}{2\rho^2}\p_t -\frac{\Delta}{2\rho^2}\p_r +\frac{a}{2\rho^2}\p_\phi, \\
\sqrt{2} (r+ia\cos\theta) m = ia \sin \theta \p_t +\p_\theta +\frac{i}{\sin \theta}\p_\phi, \\
{\bf k}^\mu{\bf l}_\mu=-1 = - m_\mu \bar{m}^\mu.  
\eeq
For the type {\bf D} metric the only nontrivial components of the Weyl tensor have boost weight zero and can be  expressed via the 
complex $\Psi_2$ invariant \cite{petr,exac}
\be\label{psi_2_Kerr}
\Psi_2\equiv -C_{\mu\nu\lambda\rho} {\bf l}^\mu m^\nu \bar{m}^\lambda {\bf k}^\rho  = \frac{M}{(r+ia\cos \theta)^{3}}.
\ee

\subsection{$r=r_0=const$ surface}
A natural candidate for the fluid hypersurface is the $r=r_0=const$ hypersurface $\Sigma$  with the  unit normal 
\be
n^\mu \p_\mu  = \frac{\sqrt{\Delta}}{\rho}  \p_r. 
\ee                                                                                                                           
The normal vector becomes null when $\Delta(r)=0$, what corresponds to the  pair of BH horizons  
\be
\Delta(r) = 0 = (r-r_+)(r-r_-),\;\;\; r_{\pm} = M\pm \sqrt{M^2-a^2}. 
\ee
The mean curvature  of $\Sigma$ 
\be\label{mean_curv_Kerr}
K = h^{ab} K_{ab}  = \nabla_\mu n^\mu = \frac{1}{\sqrt{|g|}} \p_\mu ( \sqrt{|g|} n^\mu)    = \frac{\p_r \Delta}{2\sqrt{\Delta}}(r_0) + \frac{\p_r \rho}{ \rho} \sqrt{\Delta}(r_0)
\ee
becomes large as we approach horizons. We can use $\Delta$ as small expansion parameter to describe the geometry of $\Sigma$ as it approaches the outer horizon $r=r_+$. 

\subsection{Induced metric}

The induced metric  on $\Sigma$ is 
\be\label{Kerr_ind}
ds^2 = -\left(1-\frac{2Mr}{\rho^2}\right) dt^2 - \frac{4Mar\sin^2\theta}{\rho^2}dt d\phi 
+\rho^2 d\theta^2 +\left(\frac{(r^2+a^2)^2  - \Delta a^2\sin^2\theta}{\rho^2}\right) \sin^2\theta d\phi^2, 
\ee
with $r, \rho$ being evaluated at $r_0$. Fo the rest of the section we will discuss the hypersurface quantities, so in order to avoid an overusing the $0$-subscript we will assume that  $r=r_0$. 
The determinant of the induced metric 
\be
\det h_{ab} = -\rho^2 \Delta \sin^2 \theta,   
\ee
so the hypersurface indeed becomes null as $r_0\to r_+$.  We can remove the $dtd\phi$ term by choosing the {\it zero angular momentum observable} (ZAMO) coordinates 
\be\label{Kerr_coord_transform}
\phi\to \phi'=\phi +\Omega_0 t,\;\;\; t\to t'=t,\;\;\; \Omega_0 = \frac{2Mar_0}{(a^2+r_0^2)^2- \Delta a^2\sin^2\theta}.  
\ee
The metric (\ref{Kerr_ind}) becomes
 \beq\label{Kerr_induced_metric}
 ds^2 =- \frac{\Delta \rho^2 dt^2   }{(r^2+a^2)^2  - \Delta a^2\sin^2\theta}  +\rho^2 d\theta^2  +\left(\frac{(r^2+a^2)^2  - \Delta a^2\sin^2\theta}{\rho^2}\right) \sin^2\theta \left(d\phi+ t\p_\theta\Omega_0  d\theta  \right)^2.
 \eeq
   In particular the leading $\Delta$-expansion
 \beq\label{Kerr_induced_expansion}
 h_{tt}  =  - \frac{\Delta \rho^2 }{(r^2+a^2)^2 } +\cO(\Delta^2),\;\;\; h_{t\phi} = h_{t\theta} =0,\;\; 
 h_{\phi\phi} =\frac{(r^2+a^2)^2 \sin^2\theta  }{\rho^2}  +\cO(\Delta), \\
 h_{\phi\theta} = \frac{a^3\Delta t \sin 2\theta  \sin^2\theta  }{\rho^2 (r^2+a^2)} + \cO(\Delta^2),\;\;\; h_{\theta\theta} =  \rho^2 + \cO(\Delta^2).
 \eeq

\subsection{Extrinsic curvature}
The extrinsic curvature on the hypersurface is defined via the Lie derivative 
\be
\cl_n g_{\mu\nu} = n^\sigma \p_\sigma  g_{\mu\nu}   + \p_\mu n^\sigma g_{\sigma\nu} + \p_\nu n^\sigma g_{\mu \sigma}.
\ee
Since our normal vector has only $r$-component and the Kerr metric (\ref{Kerr_BL}) has $g_{rt}=g_{r\phi }=g_{r\theta}=0$ the Lie derivative simplifies into
\be
2K_{ab}dx^adx^b = \cl_n ds^2|_{\Sigma} = n^\mu \p_\mu ds^2|_{\Sigma} =    \frac{\sqrt{\Delta}}{\rho}  \p_r ds^2.
\ee
For our hypersurface after the coordinate transform (\ref{Kerr_coord_transform}) we have the following extrinsic curvature components 
\beq
K_{tt} =   -\frac{ \rho \p_r \Delta \sqrt{\Delta} }{2(r^2+a^2)^2 }  +\cO(\Delta^{3/2}),\;\;\; K_{\theta\phi} =  \cO(\Delta^{3/2}),\;\;\;  K_{t\theta}=\cO(\Delta^{3/2}), \\
K_{t\phi} = r a\sqrt{\Delta}  \frac{\sin^2\theta}{\rho^3}   \left[     1+ \frac{ \rho^2 \p_r\Delta } {2r  (a^2+r^2) } \right]  +\cO(\Delta^{3/2}),  \;\;\;
K_{\theta\theta} =    \frac{r\sqrt{\Delta}}{\rho}  +\cO(\Delta^{3/2}),  \\
K_{\phi\phi} =   \frac{\sqrt{\Delta}\sin^2\theta }{2\rho^5}  \left(  2r(r^2+a^2)  [ 2\rho^2 -(r^2+a^2)]             - a^2 \rho^2\p_r\Delta  \sin^2\theta   \right)  +\cO(\Delta^{3/2}).
\eeq

 \subsection{Null frame}
 The fluid/gravity correspondence  require algebraic speciality with respect to  the null frame constructed from the hypersurface data $n,T$. In practice we often have a null frame of maximal speciality written for the full bulk metric. Given a null frame we can  construct the  unit vector
\be
 n = \frac{{\bf k} -{\bf l}}{\sqrt{2}}, 
\ee
but it can fail to be hypersurface orthogonal, i.e. obey 
\be
n_{[\mu} \p_\nu n_{\lambda]}=0. 
\ee 
Moreover, we need $\Sigma$ to have large mean curvature $K  = \nabla_\mu n^\mu $, so there is no canonical fluid hypersurface for a generic algebraically special geometry. 
Fortunately for  the Kerr geometry we have a family of $r=const$ surfaces that approach the horizon and serve as natural candidates for $\Sigma$. For these hypersurface the mean curvature (\ref{mean_curv_Kerr}) becomes large as we approach the horizon, 
but they do not preserve the full type {\bf D} speciality.   Let us figure out how much  speciality can we preserve on $\Sigma$.  

The  null frame (\ref{type_D_frame})  on  $\Sigma$ changes under the coordinate transformation (\ref{Kerr_coord_transform}) to become
\beq
{\bf k}\to {\bf k}=\frac{r^2+a^2}{\Delta}\p_t +\p_r  - \frac{a \rho^2}{(a^2+r^2)^2- \Delta a^2\sin^2\theta}\p_\phi,\\
{\bf  l} \to {\bf l} =\frac{r^2+a^2}{2\rho^2}\p_t -\frac{\Delta}{2\rho^2}\p_r  - \frac12\frac{\Delta}{(a^2+r^2)^2- \Delta a^2\sin^2\theta}\p_\phi.
\eeq
Moreover we can further rescale vectors ${\bf k} \to \frac{\sqrt{\Delta}}{\sqrt{2}\rho} {\bf k}$ and ${\bf l} \to \frac{\sqrt{2}\rho}{\sqrt{\Delta}}  {\bf l}$ so that 
 \beq\label{null_Kerr_delta}
 {\bf k}=\frac{T+n}{\sqrt{2}} - \frac{a \sqrt{\Delta} \rho }{\sqrt{2}(r^2+a^2)^2}\p_\phi +\cO( \Delta^{3/2}) \p_\phi+ \cO(\Delta)T,\\
  {\bf l} =\frac{T-n}{\sqrt{2}} - \frac{a \sqrt{\Delta} \rho }{\sqrt{2}(r^2+a^2)^2}\p_\phi+\cO( \Delta^{3/2}) \p_\phi+\cO(\Delta)T,
\eeq
with unit normal  to $\Sigma$ and the  time-foliation normal for the induced metric  (\ref{Kerr_induced_metric})  on $\Sigma$ being
\be
n = \frac{\sqrt{\Delta}}{\rho} \p_r,\;\;\; T = \frac{\sqrt{(r^2+a^2)^2-\Delta a^2\sin^2\theta}}{\rho\sqrt{\Delta}}\p_t  =  \frac{r^2+a^2}{\rho\sqrt{\Delta}}\p_t +\cO(\sqrt{\Delta}). 
\ee
The difference between the $\frac{T\pm n}{\sqrt{2}}$ null frame and the canonical type {\bf D} null frame (\ref{null_Kerr_delta}) can be schematically  written as $\sqrt{\Delta} f^i m_i$. The possible corrections to the algebraic constraints from 
the different weight components have the following schematic form
\beq\label{null_frame_corrections}
\delta  C_{{\bf k}i{\bf k}j} \sim \delta  C_{{\bf l}i{\bf l}j} = 2\sqrt{\Delta} f^k  C_{{\bf l}ikj} + \Delta f^k f^l  C_{ki lj} = \cO(\Delta),    \\
 \delta C_{{\bf k}ijk}  \sim  \delta C_{{\bf l}ijk} =\sqrt{\Delta} f^l  C_{lijk} = \cO(\sqrt{\Delta}). 
\eeq
In the second equality we used the type {\bf D} constraints and the near-horizon scaling of the boost weight zero components (\ref{psi_2_Kerr}).  Using  (\ref{riemann_on_surface}) we can estimate the  leading order expressions  for the Weyl tensor components in terms of the hypersurface data to be 
 \be
  C_{{\bf l}i{\bf l}j} \sim  C_{{\bf k}i{\bf k}j} =   \cO(\Delta^0), \;\;\;\; C_{{\bf l}ijk} \sim    C_{{\bf k}ijk}  = \cO(\sqrt{\Delta}). 
 \ee
Thus the corrections  (\ref{null_frame_corrections}) does not affect only the leading order for the type {\bf I}$_i$ constraints. So effectively we can say that the choice of $\Sigma$ breaks the Kerr geometry speciality from the type {\bf D} to the subtype {\bf I}$_i$, which has 
\be\label{type_I_i_slow}
  C_{{\bf l}i{\bf l}j} = C_{{\bf k}i{\bf k}j} =0.
\ee

\section{Slowly rotating black hole}
We can consider a simplified version of the Kerr geometry (\ref{Kerr_BL}) with $a/M\ll1$, so we can drop the higher orders in $a$.  Such toy version contains almost all features of the full Kerr geometry while being very easy to analyze. 

\subsection{Geometry}
The Kerr metric (\ref{Kerr_BL}) at leading order in $a$  becomes 
\beq\label{Kerr_BL_slow}
ds^2 = -f^2 dt^2 - \frac{4Ma\sin^2\theta}{r}dt d\phi +f^{-2} dr^2
+r^2 (d\theta^2 +\sin^2\theta d\phi^2 )+\cO(a^2), \;\; f^2=1-\frac{2M}{r}.
\eeq
The  $r=r_0$ hypersurface has the following geometric data 
\beq
 n =f_0\p_r, \;\; T = f^{-1}_0 \p_t,\;\; ds^2 =  -f_0^2 dt^2 - \frac{4Ma\sin^2\theta}{r_0}dt d\phi +r_0^2 (d\theta^2 +\sin^2\theta d\phi^2 )+\cO(a^2), \\
 K_{ab}dx^adx^b =   -f_0^2 \p_r f_0 dt^2 + \frac{2Ma f_0 \sin^2\theta}{r^2_0}dt d\phi +r_0f_0  (d\theta^2 +\sin^2\theta d\phi^2 )+\cO(a^2). 
\eeq
Using the coordinate transformation 
\be\label{coord_transf_slow}
\phi \to \phi + \frac{2Ma}{r_0^3} t 
\ee
we can remove the $dtd\phi$ term, so the  geometry simplifies into
\beq\label{hyp_geometry_kerr_slow}
 ds^2 =  -f_0^2 dt^2 +r_0^2 (d\theta^2+\sin^2 \theta d\phi^2 )+\cO(a^2), \\
 K_{ab}dx^adx^b =   -f_0^2 \p_r f_0 dt^2 + \frac{6Ma f_0 \sin^2\theta}{r^2_0}dt d\phi +r_0f_0   (d\theta^2+\sin^2 \theta d\phi^2 )+\cO(a^2).
\eeq
The mean curvature 
\be
K = K_{ab}h^{ab} = \p_r f_0 + \frac{2f_0}{r_0} = \frac{M}{f_0 r_0^2} + \frac{2f_0}{r_0}
\ee
becomes large as we approach the black hole horizon at $r=2M$ since $r_0f^2_0 = (r_0-2M)$. Moreover the leading near-horizon behavior of the mean curvature is related to the Hawking temperature  $T_H$ for the slowly rotating Kerr solution.   
\be\label{BH_temperature}
2\pi T_H  = \frac{ -K_{tt}}{\sqrt{-g_{tt}}} \Big|_{r_0=r_H=2M} = f_0\p_r f_0 \Big|_{r_0=r_H=2M}=   \frac{M}{r_H^2} = \frac{1}{4M}
\ee
The type {\bf D} null frame (\ref{type_D_frame}) after  the coordinate transformation (\ref{coord_transf_slow}) becomes 
 \beq
{\bf k} =f_0^{-2}\left(\p_t - \frac{2Ma}{r_0^3}  \p_\phi\right) +\p_r +\frac{a}{r_0^2f_0^2}\p_\phi +\cO(a^2),\\
{\bf l} =\frac{1}{2}\left(\p_t - \frac{2Ma}{r_0^3}  \p_\phi\right) -\frac{f_0^2}{2}\p_r +\frac{a}{2r_0^2}\p_\phi+\cO(a^2), 
\eeq
We can freely rescale our null vectors  ${\bf k} \to f_0 {\bf k}/\sqrt{2},\;\; {\bf l} \to \sqrt{2}  {\bf l}/f_0$ without changing the null frame  
\beq
{\bf k} =\frac{1}{f_0\sqrt{2}}\p_t +\frac{f_0}{\sqrt{2}}\p_r +\frac{af_0}{\sqrt{2}r_0^2}\p_\phi +\cO(a^2) = \frac{1}{\sqrt{2}} ( T+n) +\frac{af_0}{\sqrt{2} r_0^2}\p_\phi  +\cO(a^2),\\
{\bf  l} =\frac{1}{f_0\sqrt{2}}\p_t -\frac{f_0}{\sqrt{2}}\p_r +\frac{af_0}{\sqrt{2}r_0^2}\p_\phi  +\cO(a^2) = \frac{1}{\sqrt{2}} ( T-n)  +\frac{af_0}{\sqrt{2} r_0^2}\p_\phi +\cO(a^2).
\eeq
to arrive into canonical form (\ref{canonical_frame}), used for the dual fluid construction. The additional contribution along the $\p_\phi$ breaks the type {\bf D} to type {\bf I}$_i$
(\ref{type_I_i_slow}). The first relation  in (\ref{type_I_i_slow}) is the type {\bf I} relation which we can use to describe the dual fluid. The second relation  in (\ref{type_I_i_slow}) imposes additional relations for the dual fluid, which we will describe in fluid variables later. 

\subsection{Dual fluid}
Our geometry  (\ref{hyp_geometry_kerr_slow}) can be written in terms of $\tau =\lambda^{-2}t $ and $\lambda^2 =f^2_0  = \cO(r_0-2M)$ 
\be\label{hyp_geometry_slow}
ds^2 = -  \lambda^{-2} d\tau^2 + \gamma_{ij}(x) dx^i dx^j, \;\;\;\; K = \cO(\lambda^{-1}),  
\ee
to simplify the large mean curvature expansion analysis. The type {\bf I} constraint (\ref{type_I}) for metric  (\ref{hyp_geometry_slow}) simplifies into 
\be\label{type_I_slow_fluid}
2C_{{\bf l} i  {\bf l} j } =  -\lambda  D_i K_{j\t} - \lambda D_j K_{i\t} +2 \lambda\p_\t  K_{ij}  - R^{(p)}_{ij}+  (\gamma^{kl}K_{kl}-2\lambda^2  K_{tt})K_{ij} - K_{ik}K^k_{~j}  +2\lambda^2 K_{i\t}K_{j\t}  
\ee
with $R^{(p)}_{ij}, D_i$ being Ricci tensor and  covariant derivative with respect to the spatial metric $\gamma_{ij}$. The large mean curvature  $K = \cO(\lambda^{-1})$ motivates to search for a  perturbative in   $\lambda$
solution to the type {\bf I} constraint (\ref{type_I_slow_fluid}) using
\be
\lambda K_{i\t} = \sum_{k=0}^\infty \lambda^k  K_{i\t}^{(k)},\;\;\;\; K_{ij} = \sum_{k=1}^\infty \lambda^k  K_{ij}^{(k)},\;\;\; \lambda^3 K_{\t\t} =  \sum_{k=0}^\infty \lambda^k  K_{\t\t}^{(k)}.
\ee
The solution for   $K_{ij}$  is 
\beq\label{K_ij_slow}
2K^{(0)}_{\t\t} K^{(1)}_{ij} =  -D_i K^{(0)}_{j\t} -D_jK^{(0)}_{i\t} -  R^{(p)}_{ij} +2 K^{(0)}_{i\t}K^{(0)}_{j\t}. 
\eeq
The covariant conservation equation (\ref{covariant_cons}) for our geometry (\ref{hyp_geometry_slow}) becomes 
\be\label{t_covariant_slow}
 D^i K_{i\t} -  \gamma^{ij}\p_\t  K_{ij}  =0, 
\ee
\be\label{i_covariant_slow}
  -\lambda^2  \p_\t K_{i\t}  +D^j K_{ij} -  \p_i ( \gamma^{kl} K_{kl} )    +\lambda^{2} \p_i K_{\t\t}=0.  
  \ee
The  equation  (\ref{i_covariant_slow}) at leading orders   $\cO(\lambda^{-1}, \lambda^{0})$ order is 
\be
 \p_i  K^{(0)}_{\t\t}=0, \;\;\; \p_i  K^{(1)}_{\t\t}=0.
\ee
which is the statement that the BH temperature  (\ref{BH_temperature}) is the same at each point on the horizon. The next order equation is 
\be\label{i_covariant_slow_exp}  
    - \p_\t K^{(0)}_{i\t}  +D^j K^{(1)}_{ij} - \p_i ( \gamma^{kl} K^{(1)}_{kl} )    +\p_i K^{(1)}_{\t\t}=0. 
\ee
Let us define    the fluid velocity $v_i$
\be
v_i (x,\t) \equiv K^{(0)}_{i\t},    
\ee
viscosity  $\eta$ and pressure  $p$
\be
\eta^{-1}(x,\t) \equiv -  K^{(0)}_{\t\t},\;\;\;  \eta > 0,\;\;\; p(x,\t)  \equiv  -K^{(2)}_{\t\t}.    
\ee 
The solution (\ref{K_ij_slow}) in the new notation is of the form 
\be
 2 K^{(1)}_{ij} = -2\eta   v_{i}v_{j} +\eta  (D_i v_j +D_j v_i)   +  \rho R^{(p)}_{ij}.   
\ee
The time component of the constraint equation (\ref{t_covariant_slow}) becomes the incompressibility condition for the dual fluid
\be
 D_i v^i  = 0.  
\ee
while the other component  (\ref{i_covariant_slow_exp})  
\be
  \p_\t v^i   +  \eta  v_j  D^j v^i     -  \frac12   \eta D_j  (D^i v^j +D^j v^i)        +\p^i \left[p  +\frac14  \eta  R^{(p)} - \eta v^2  \right]=0, 
\ee
becomes the Navier-Stokes equation  in curved space
\be\label{toy_NS}
  \p_\t v^i   +   v_j  D^j v^i     -  \frac12   \eta D_j  (D^i v^j +D^j v^i)    +\p^i p=0, 
\ee
if we redefine pressure and velocity  
\be
v^i \to \eta^{-1} v_i, \;\;\;\eta p    +\frac14 \eta^2 R^{(p)} - v^2 \to p.
\ee
 The equation (\ref{toy_NS}) describes the shear perturbation for the Schwarzschild metric \cite{Bredberg:2011xw}, so it is not surprising that  it describes the dual fluid for the slowly-rotating Kerr black hole.

\subsection{Type {\bf I}$_{i}$ fluid as a Killing flow}
 Our geometry  (\ref{Kerr_BL_slow}) and the fluid hypersurface  admits an additional vanishing Weyl tensor component (\ref{type_I_i_slow}).  We already have the same number of fluid variables and fluid equations  so any additional constraints would restrict the generality of the fluid solution. The additional constraint can be written 
 \be
  0=C_{{\bf k}i{\bf k}j}-C_{{\bf l}i{\bf l}j} = C_{n i Tj } +C_{Tinj} =  R_{n i Tj } +R_{Tinj}  - \frac{2}{p} \gamma_{ij} R_{nT}, 
 \ee  
so we can use the Riemann tensor instead. The leading order expression  in fluid variables    
\be\label{type_I_i_fluid}
R_{n i Tj } +R_{Tinj} =0 \Rightarrow  D_i v_j +D_j v_i=\cl_v \gamma_{ij}=0. 
\ee
The equation (\ref{type_I_i_fluid}) is a Killing equation, so the solution for a fluid velocity can be written as 
\be\label{Killing_sol}
v_i (x,\t) =f(\t) k_i (x),
\ee 
with $k_i(x)$ being a Killing vector for the spatial metric $\gamma_{ij}$.   The substitution of (\ref{Killing_sol}) into (\ref{toy_NS}) lead to the equation
\be
\p_\t v^i =  \p^i \left(\frac12  v^2-p\right).
\ee 
We can integrate over the fluid on compact spatial section of the horizon 
 \be
 \p_\t  \int \sqrt{\gamma}  v^2 = \p_\t (f^2) \int k^2 \sqrt{\gamma} =0  
 \ee
to conclude that $\p_\tau f=0$. So we can summarize that Killing-based solution (\ref{Killing_sol}) solves the NS equation  (\ref{toy_NS}) with  constant $f(\tau)$, that is imposed by compactness of the Kerr horizon. 
The solution is 
\be\label{toy_NS_solution}
v_i =k_i(x),\;\;\; P = \frac12  k_i k^i (x) +\hbox{const}.  
\ee
The solution  (\ref{toy_NS_solution}) is generic for any type {\bf I}$_i$ metric with a hypersurface geometry (\ref{hyp_geometry_slow}) and large mean curvature. Our slowly rotating Kerr geometry (\ref{hyp_geometry_kerr_slow}) obeys these properties  so we can compare extrinsic curvature  from (\ref{hyp_geometry_kerr_slow}) and the  NS solution   (\ref{toy_NS_solution}). The  dual fluid velocity 
\be\label{slow_kerr_v_sol}
v^i \p_i  = \eta^{-1}\gamma^{ij}K_{\tau j} \p_i = \frac{3a }{r^2_0} \p_\phi  
\ee
is  $\tau$-independent  Killing vector on the round two sphere. Indeed our result for the hypersurface geometry (\ref{hyp_geometry_slow}) match the generic solution for toy example (\ref{toy_NS_solution}).

Our toy example illustrates an  feature, that are common for all type {\bf D} dual fluids: Fluid velocity is closely related to the    Killing vectors of the metric. The NS equation in curved space (\ref{toy_NS}) always admits 
solutions from Killing vectors (\ref{toy_NS_solution}). On the gravity  side it is conjectured that the type {\bf D} metrics have some number of Killing vectors and fluid/gravity mapping lead to Killing equation as one of the additional constraints present for type {\bf D} metrics.

\section{Dual fluid for rotating black hole}
The most studied fluid system is the incompressible Navier-Stokes  (NS) equation, so we want to use additional physical assumptions to write  our fluid system in the form similar to the NS equation. 
In particular we will use the near-horizon expansion with small parameter $\lambda$ to control the relative size of various terms and simplify our equations.  

\subsection{Geometry}
The induced metric on the hypersurface is of the form (\ref{Kerr_induced_expansion})
\be
ds^2 = h_{ab} dx^a dx^b =  -\lambda^2 h(x) dt^2 +\gamma_{ij}(x,\lambda ^2 t) dx^i dx^j. 
\ee
where we we identify $\lambda^2 r^2 = \Delta$. The smallness $h_{tt}=\cO(\lambda^2)$   is a generic feature when  $\Sigma$ approaches the black hole horizon and becomes null. 
The second property,  $\p_t \gamma_{ij} = \cO(\lambda^2)$, is related to the fact that all point on the black hole horizon move with the same angular velocity $\Omega_H$.  Let us do a time redefinition  $\tau =\lambda^2 t $  so the metric takes the form 
\be\label{rot_bh_metric}
ds^2 = - h(x)\frac{ d\tau^2}{\lambda^2}  +\gamma_{ij} (x,\tau) dx^i dx^j,\;\;\; \p_\tau \gamma_{ij} = \cO(\lambda^0).  
\ee
We chose $\tau=const$ as a time foliation so the unit normal is of the form 
\be
T = \lambda h^{-1/2} \p_\tau. 
\ee
Using our explicit metric ansatz  (\ref{rot_bh_metric}) we can rewrite the type {\bf I} constraint (\ref{type_I}) 
\beq
2C_{{\bf l} i  {\bf l} j } = - \lambda h^{-1} D_i ( h^{1/2}K_{j\t}) - \lambda h^{-1}D_j( h^{1/2}K_{i\t} ) +2 \lambda h^{-1/2}\p_\t  K_{ij} + h^{-3/2}\lambda^3 \p_\t \gamma_{ij}  K_{\t\t} \\
 -\frac12\lambda  h^{-1/2} \p_\t \gamma_{il} \gamma^{kl} K_{kj}  -\frac12 \lambda h^{-1/2} \p_\t \gamma_{jl} \gamma^{kl} K_{ik}+
D_iD_j \log h +\frac12 \p_i \log h \p_j \log h\\ - R^{(p)}_{ij}+  (K-\lambda^2 h^{-1} K_{\tau\tau})K_{ij} - K_{ik}K^k_{~j}  +2\lambda^2 h^{-1}K_{i\tau}K_{j\tau} +\cO(\lambda^2).
\eeq
\subsection{Near-horizon expansion}
Using the large mean curvature  $K = \cO(\lambda^{-1})$ of a hypersurface and the  $\lambda$-expansion
\be
\lambda K_{i\t} = \sum_{k=0}^\infty \lambda^k  K_{i\t}^{(k)},\;\;\;\; K_{ij} = \sum_{k=1}^\infty \lambda^k  K_{ij}^{(k)},\;\;\; \lambda^3 K_{\t\t} =  \sum_{k=0}^\infty \lambda^k  K_{\t\t}^{(k)}, 
\ee
we can solve for $K_{ij}$ 
\beq\label{K_ij}
 2K^{(0)}_{\tau\tau} K^{(1)}_{ij}  = -  D_i ( h^{1/2}K^{(0)}_{j\t}) -  D_j( h^{1/2}K^{(0)}_{i\t} )+ h^{-1/2} \p_\t \gamma_{ij}  K^{(0)}_{\t\t} +\\
h D_iD_j \log h +\frac12 h\p_i \log h \p_j \log h -h R^{(p)}_{ij} +2 K^{(0)}_{i\tau}K^{(0)}_{j\tau}. 
\eeq
The covariant conservation (\ref{covariant_cons})  for our metric becomes 
\be\label{cov_cons_kerr}
   2 h^{-1}  \lambda D^i ( h^{1/2}K_{i\t})  - h^{-3/2}\lambda^3 \gamma^{ij} \p_\t \gamma_{ij}  K_{\t\t} -2 \lambda h^{-1/2} \gamma^{ij}\p_\t  K_{ij} + \lambda h^{-1/2}K^{ij}  \p_\t \gamma_{ij} =0 
\ee
and 
\be
  -\lambda^2  \p_\t( h^{-1/2}K_{\t i})  +D^j (\sqrt{h}K_{ij}) - \sqrt{h} \p_i ( \gamma^{kl} K_{kl} )   - \frac12 h^{-1/2}  \lambda^2 \p_\t \log  \gamma K_{\t i}   +\lambda^{-2} \p_i (h^{-1/2} K_{\t\t})=0. 
  \ee
 Our equation at leading $\cO(\lambda^{-1},\lambda^0)$ orders is 
\be
 \p^i (h^{-1/2} K^{(0)}_{\t\t})=0, \;\;\; \p^i (h^{-1/2} K^{(1)}_{\t\t})=0,  
\ee
which is the statement that the BH temperature is the same at each point on the horizon. The next order is 
\be\label{i_cov_cons_exp}  
    - \p_\t( h^{-1/2}K^{(0)}_{\t i})  +D^j (\sqrt{h}K^{(1)}_{ij}) - \sqrt{h} \p_i ( \gamma^{kl} K^{(1)}_{kl} )   - \frac12 h^{-1/2}  \p_\t \log  \gamma K^{(0)}_{\t i}   +\p_i (h^{-1/2} K^{(2)}_{\t\t})=0.  
\ee
Similarly to the slow rotating case let us  introduce  the fluid velocity $v_i$ as 
\be
v_i (x,\t) \equiv h^{1/2}  K^{(0)}_{i\t},     
\ee
viscosity  $\rho$ and pressure  $p$ 
\be
\eta^{-1}(x,\t) \equiv -h^{-1/2}K^{(0)}_{\t\t} ,\;\;\;  p(x,\t)  \equiv  -h^{-1/2}K^{(2)}_{\t\t}.
\ee 
The solution (\ref{K_ij}) in the new notation is of the form 
\be
 2 \sqrt{h}K^{(1)}_{ij} = -2\eta h^{-1}  v_{i}v_{j} +\eta  (D_i v_j +D_j v_i)  -\eta h D_iD_j \log h -\frac12 \eta h \p_i \log h \p_j \log h+ \eta hR^{(p)}_{ij}    +   \p_\t \gamma_{ij}.  
\ee
The time-component of the constraint equations (\ref{cov_cons_kerr}) becomes the incompressibility equation 
\be
\eta D_i v^i  + \frac12  \p_t \log \gamma =0, 
\ee
while the spatial component  (\ref{i_cov_cons_exp}) becomes the NS-like equation
\beq\label{NS_kerr}
     h^{-1}\p_\t v_i      + \eta h^{-1}  v^{j} D_j v_{i}   -\eta h^{-2}  v_{i}v_{j} \p^j h -\frac12 \eta  D^j(D_i v_j +D_j v_i +\eta^{-1}   \p_\t \gamma_{ij}   )  +\frac12 \eta h^{-2} v^2 \p_i h\\
+  \p_i (  p + \frac14 \eta hR^{(p+1)} + \frac14\eta  D^2 h   -\eta  h^{-1}  v^2) =0.
\eeq
The equation (\ref{NS_kerr}) is not the NS equation in curved space, since there are several additional terms.  So, unfortunately, the fluid gravity map for the Kerr black hole cannot be used to generate a 
new solution to the usual NS system. However, we should notice that the   (\ref{NS_kerr})  depends on the coordinate choice for the  hypersurface metric (\ref{rot_bh_metric}). For example we can choose the Gaussian normal coordinates, where $h(x)=1$ at the price of more complicated $\tau$-dependence  for the spatial metric $\gamma_{ij}$. Thus, the reasonable questions about the dual fluid should be formulated in invariant form. 
In the remaining part of the paper we will describe two features of the dual fluid (Killing-based solution and linearized stability), that are independent on particular coordinate choice and are also present for the usual NS equation.

\subsection{The type {\bf I}$_{i}$ fluid as a Killing flow}
The additional constraint for type  {\bf I}$_{i}$  geometries  $C_{{\bf k} i {\bf k} j}=0$ can be written using Riemann tensor 
\be\label{type_I_i_fluid_gen}
R_{n i Tj } +R_{Tinj} =0  \Rightarrow  \Sigma_{ik}\equiv D_i v_j +D_j v_i +   \eta^{-1} \p_\t \gamma_{ij}=0.   
\ee
The NS equation (\ref{NS_kerr}) can be rewritten using $\Sigma_{ij}$
\be
h^{-1}\gamma_{ij}\p_\t v^j      +\eta h^{-1}  v^{j} \Sigma_{ij}    -\eta h^{-2}  v_{i}v_{j} \p^j h -\frac12 \eta D^j\Sigma_{ij} 
+  \p_i \left(  p + \frac14\eta hR^{(p+1)} + \frac14 \eta  D^2 h   -\frac32 \eta  h^{-1}  v^2\right) =0,      
\ee
so  the additional  constraint (\ref{type_I_i_fluid_gen})  reduces  equation into 
\be
  h^{-1}\dot{ v}^i      -  \eta h^{-2}  v^{i} v^{k} \p_k h  +\p^i \left(  p + \frac14\eta hR^{(p+1)} + \frac14 \eta  D^2 h   -\frac32 \eta  h^{-1}  v^2\right)=0. 
\ee
Furthermore, algebraically special solutions of type {\bf D} conjectured to have some number of Killing vectors \cite{Coley:2007tp}.    The additional constraint (\ref{type_I_i_fluid_gen}) can  be written as a Killing equation 
\be
\p_\t \gamma_{ij} = D_j w_i +D_i w_j ,\;\;\; D_i v_j +D_j v_i +   \eta^{-1} \p_\t \gamma_{ij}    =  D_i (v_j + \eta^{-1}w_j) +D_j (v_i +\eta^{-1} w_i),  
\ee 
since  for Kerr geometry and some other examples time dependence of the metric $\gamma$ comes from a coordinate transformation $x^i \to x^i +w^i \t$ (\ref{Kerr_coord_transform}).   In case of Kerr geometry both fluid velocity $v_i$ and  $w_i$ are aligned with $\p_\phi$ bulk Killing vector so $v^k \p_k h$ term vanishes.  In particular the fluid velocity for   our explicit Kerr data: 
\be
v^i \p_i =  h^{1/2}  \gamma^{ij}K^{(0)}_{\t j}\p_i =  \frac{r^3 a}{(r^2+a^2)^3 }  \left[     1+ \frac{ \rho^2 \p_r\Delta } {2r  (a^2+r^2) } \right]  \p_\phi.
\ee
Thus the NS equation for the Kerr case is solved by the static velocity, that is almost Killing vector.  Given existence of the multiple Killing vectors for a higher dimensional type {\bf D}
geometries we can conjecture that the corresponding dual fluid solutions have similar structure of the static  almost Killing flows.

\section{Stability, Reynolds number, Turbulence.}

The most interesting phenomenon accompanying the fluid dynamics is the turbulence.  So it might be very fruitful to see what can we say about it using the  fluid/gravity correspondence. 
There are two major directions of turbulence studying: {\it developing turbulence} and {\it developed turbulence}. The first direction address questions about instabilities of the fluid flow, their emergence, dynamics and interactions.  
So in context of developing turbulence we are interesting in a transition process from smooth flow to a chaotic one. In case of developed turbulence we usually assume that  the flow is sufficiently chaotic, so we can study 
interesting relations  for the statistically averaged quantities. We do not have much to say about the developed turbulence while we can use our explicit results for the Kerr fluid to study its stability properties. 

Let us use our toy example to describe the notion of stability and its features.  The NS equation on static curved background  is 
\be\label{NS_big}
\p_\t v_i  + v^j D_j v_i -\eta D^j (D_i v_j +D_j v_i)  +\p_i p=0, \;\;\;\; D_i v^i=0.\ee
For a given solution $V, P$ we can consider a small perturbation on top of it 
\be
v \to V + v ,\;\;\; p \to  P +p,
\ee
that satisfies the following linearized equation
\be\label{linearized_NS}
\p_\t v_i  + V^j D_j v_i + v^j D_j V_i -\eta D^j (D_i v_j +D_j v_i) +\p_i p=0, \;\;\;\; D_i v^i=0.
\ee
Our equations (\ref{linearized_NS}) are linear so it is useful to expand $v(x,\t)$ in Fourier modes. For a given mode $\hat{v} (\omega, k)$ its time evolution $e^{-i\omega \t}$ is governed by the frequency $\omega$. 
In general $\omega$ is a complex number, so given a positive imaginary part we may have a perturbations that grow with time. It is natural do define solution $V,P$ to the NS system  (\ref{NS_big}) as a stable solution if 
the linearized equation  (\ref{linearized_NS}) has no Fourier modes with negative imaginary part, i.e. all linearized solutions decay with time. A complete analysis for linearized stability using mode expansions can only be done  for simplest fluid solutions, so we need use some other methods for generic flows.  

The powerful approach to linearized stability is based on {\it energy balance equation}\cite{TheoLib:GEN15}. Let us take the linearized equation (\ref{linearized_NS}),  multiply it by $v_i$ and integrate over the time slice of the horizon
\be\label{energy_balance}
\frac12 \p_\tau \int \sqrt{\gamma} v^2 =  -\frac{\eta}{2}  \int \sqrt{\gamma} ( D_jv_i+D_i v_j)^2  - \int \sqrt{\gamma} v^i v^k D_k V_i. 
\ee
The lefthandside measures time dependence for the kinetic energy of the fluid perturbation while the rightahndside describes the energy dissipation by viscosity term and energy flow from background solution. 
The relative size for the two terms in  (\ref{energy_balance}) can be described in terms of dimensionless combination 
\be
Re = \frac{Lv}{\eta}, 
\ee
so that stability requires 
\be
Re<Re_c. 
\ee
The critical value is determined by minimization 
\be
Re_c \equiv \min \frac{ \frac12 \int \sqrt{\gamma} ( D_jv_i+D_i v_j)^2}{-\int \sqrt{\gamma} v^i v^k D_k V_i}  
\ee
over divergence free velocities. This bound was first proposed by O.Reynolds \cite{reynolds}.  Existence of the minimum is guarantied because of the same (quadratic) velocity dependence for both functionals.  
 Moreover,  the  contribution to energy balance from the quadratic term, that we dropped in linearized analysis,  is a total derivative 
\be
v^i v^j D_j v_i  =  \frac12 D_j( v^j v^2) 
\ee  
So this extra term does not affect our analysis.  Let us apply the energy balance analysis to the toy solution and to  generic rotational black hole solution. 

Reynolds number for the toy solution (\ref{slow_kerr_v_sol}) 
\be
Re = \frac{v L}{\eta} = 3\frac{a}{M} = \frac{3J}{M^2}.
\ee
By construction it is parametrically small for slow-rotating Kerr, what is usually sufficient for stable fluid.  Nevertheless let us evaluate the $Re_c$. 
Our toy case fluid solution  (\ref{toy_NS_solution})  has   trivial energy flow from background solution  
\be
\int \sqrt{\gamma} v^i v^k D_k V_i = \frac12 \int \sqrt{\gamma} v^i v^k ( D_k V_i +D_i V_k) =0,
\ee
since background fluid velocity obeys Killing equation.  So the Killing flow solution (\ref{toy_NS_solution})  for slowly rotating black hole has $Re_c=\infty$ and it is always stable. 

For the generic Kerr we have the following linearized system 
\be
D_i v^i=0,
\ee
\beq
h^{-1}\gamma_{ij}\p_\t v^j      + \eta h^{-1}  v^{j} \Sigma_{ij}  +\eta h^{-1}  V^{j} (D_i v_j +D_j v_i )    -\eta h^{-2}  V_{i}v_{j} \p^j h  -\eta h^{-2}  v_{i}V_{j} \p^j h\\
 -\frac12 \eta  D^j(D_i v_j +D_j v_i)+  \p_i    p =0.  
\eeq
The extra terms are similar to the ones described in \cite{Wu:2015pzg} and describe various physical effects like Coriolis force.

The energy balance equation is
\be
\gamma^{-1/2} h^{-1}  \p_\tau  (\gamma^{1/2} v^2) = -\frac12 \eta (D_i v_j +D_j v_i)^2 -  \eta h^{-1}  v^iv^{j} \Sigma_{ij}     + \eta h^{-2}  v^2V_{j} \p^j h    -  D_i (v^i   p  +....).    
\ee
The type {\bf I}$_i$ condition $\Sigma_{ij}=0$ and bulk Killing symmetry $V^k \p_k h=0$ simplify energy balance into 
\be
 \p_\tau  \int h^{-1} v^2 \sqrt{\gamma} = -   \frac12\eta \int   (D_i v_j +D_j v_i)^2     \sqrt{\gamma},
\ee
what makes the fluid, dual to rotating black hole, stable.

\section{Summary of the results }

We described the generalization of the algebraically special  fluid/gravity correspondence, which includes rotating black holes in four and higher dimensions. On a gravity side rotating black holes are type {\bf D}
algebraically special while the  duality construction requires to choose the null frame with  smaller speciality. The reduced speciality is of type {\bf I}$_{i}$, which is still bigger then  the minimal type {\bf I} required by the duality. 
Thus the dual fluid obeys additional equations, which are very similar to the Killing equations. In case of slowly-rotating Kerr fluid velocity is e  the Killing vector for the $S^2$, which is the constant time slice of the horizon.
For a generic Kerr    solution fluid velocity is closely related to the bulk Killing vector  and this pattern seems to have a generalization to higher dimensions.  

The dual fluid system for slowly-rotating Kerr is a simple generalization of the  incompressible Navier-Stokes equation to the curved background. While the fluid equations for the generic Kerr solution contain 
various additional terms comparing to the  Navier-Stokes equation. Moreover, the structure of additional terms depends on a particular choice of the hypersurface metric.   However, there are interesting common features  
between Kerr-dual fluid and normal incompressible fluid.  One of them is that the both equations admit Killing-like solutions.  The second feature is the energy balance equation for the Killing-flow  background, which can be used to show that such flows are stable.  

\section*{Acknowledgements}
 Author is grateful to A. Strominger for useful discussions on early stages of this work. This work was supported in part by  DOE grant DE-SC0011632 and  the Sherman Fairchild scholarship.

\bibliography{Kerr_fluid_ref}{}

\providecommand{\href}[2]{#2}\begingroup\raggedright\begin{thebibliography}{10}

\bibitem{Price:1986yy}
R.~H. Price and K.~S. Thorne, ``{Membrane Viewpoint on Black Holes: Properties
  and Evolution of the Stretched Horizon},''
\href{http://dx.doi.org/10.1103/PhysRevD.33.915}{{\em Phys. Rev.} {\bfseries
  D33} (1986) 915--941}.
%%CITATION = PHRVA,D33,915;%%.

\bibitem{Bhattacharyya:2008jc}
S.~Bhattacharyya, V.~E. Hubeny, S.~Minwalla, and M.~Rangamani, ``{Nonlinear
  Fluid Dynamics from Gravity},''
  \href{http://dx.doi.org/10.1088/1126-6708/2008/02/045}{{\em JHEP} {\bfseries
  02} (2008) 045},
\href{http://arxiv.org/abs/0712.2456}{{\ttfamily arXiv:0712.2456 [hep-th]}}.
%%CITATION = ARXIV:0712.2456;%%.

\bibitem{Bhattacharyya:2008kq}
S.~Bhattacharyya, S.~Minwalla, and S.~R. Wadia, ``{The Incompressible
  Non-Relativistic Navier-Stokes Equation from Gravity},''
  \href{http://dx.doi.org/10.1088/1126-6708/2009/08/059}{{\em JHEP} {\bfseries
  08} (2009) 059},
\href{http://arxiv.org/abs/0810.1545}{{\ttfamily arXiv:0810.1545 [hep-th]}}.
%%CITATION = ARXIV:0810.1545;%%.

\bibitem{bkls}
I.~Bredberg, C.~Keeler, V.~Lysov, and A.~Strominger, ``{Wilsonian Approach to
  Fluid/Gravity Duality},''
\href{http://arxiv.org/abs/1006.1902}{{\ttfamily arXiv:1006.1902 [hep-th]}}.
%%CITATION = 1006.1902;%%.

\bibitem{Bredberg:2011jq}
I.~Bredberg, C.~Keeler, V.~Lysov, and A.~Strominger, ``{From Navier-Stokes To
  Einstein},''
\href{http://arxiv.org/abs/1101.2451}{{\ttfamily arXiv:1101.2451 [hep-th]}}.
%%CITATION = 1101.2451;%%.

\bibitem{Son:2007vk}
D.~T. Son and A.~O. Starinets, ``{Viscosity, Black Holes, and Quantum Field
  Theory},''
  \href{http://dx.doi.org/10.1146/annurev.nucl.57.090506.123120}{{\em Ann. Rev.
  Nucl. Part. Sci.} {\bfseries 57} (2007) 95--118},
\href{http://arxiv.org/abs/0704.0240}{{\ttfamily arXiv:0704.0240 [hep-th]}}.
%%CITATION = ARXIV:0704.0240;%%.

\bibitem{Baier:2007ix}
R.~Baier, P.~Romatschke, D.~T. Son, A.~O. Starinets, and M.~A. Stephanov,
  ``{Relativistic viscous hydrodynamics, conformal invariance, and
  holography},'' \href{http://dx.doi.org/10.1088/1126-6708/2008/04/100}{{\em
  JHEP} {\bfseries 04} (2008) 100},
\href{http://arxiv.org/abs/0712.2451}{{\ttfamily arXiv:0712.2451 [hep-th]}}.
%%CITATION = ARXIV:0712.2451;%%.

\bibitem{Lysov:2011xx}
V.~Lysov and A.~Strominger, ``{From Petrov-Einstein to Navier-Stokes},''
\href{http://arxiv.org/abs/1104.5502}{{\ttfamily arXiv:1104.5502 [hep-th]}}.
%%CITATION = ARXIV:1104.5502;%%.

\bibitem{Coley:2007tp}
A.~Coley, ``{Classification of the Weyl Tensor in Higher Dimensions and
  Applications},'' \href{http://dx.doi.org/10.1088/0264-9381/25/3/033001}{{\em
  Class. Quant. Grav.} {\bfseries 25} (2008) 033001},
\href{http://arxiv.org/abs/0710.1598}{{\ttfamily arXiv:0710.1598 [gr-qc]}}.
%%CITATION = ARXIV:0710.1598;%%.

\bibitem{Bhattacharyya:2007vs}
S.~Bhattacharyya, S.~Lahiri, R.~Loganayagam, and S.~Minwalla, ``{Large rotating
  AdS black holes from fluid mechanics},''
  \href{http://dx.doi.org/10.1088/1126-6708/2008/09/054}{{\em JHEP} {\bfseries
  09} (2008) 054},
\href{http://arxiv.org/abs/0708.1770}{{\ttfamily arXiv:0708.1770 [hep-th]}}.
%%CITATION = ARXIV:0708.1770;%%.

\bibitem{Bredberg:2011xw}
I.~Bredberg and A.~Strominger, ``{Black Holes as Incompressible Fluids on the
  Sphere},'' \href{http://dx.doi.org/10.1007/JHEP05(2012)043}{{\em JHEP}
  {\bfseries 05} (2012) 043},
\href{http://arxiv.org/abs/1106.3084}{{\ttfamily arXiv:1106.3084 [hep-th]}}.
%%CITATION = ARXIV:1106.3084;%%.

\bibitem{Wu:2015pzg}
C.-J. Chou, X.~Wu, Y.~Yang, and P.-H. Yuan, ``{Rotating Black Holes and
  Coriolis Effect},''
  \href{http://dx.doi.org/10.1016/j.physletb.2016.08.018}{{\em Phys. Lett.}
  {\bfseries B761} (2016) 131--135},
\href{http://arxiv.org/abs/1511.08691}{{\ttfamily arXiv:1511.08691 [hep-th]}}.
%%CITATION = ARXIV:1511.08691;%%.

\bibitem{petr}
A.~Petrov, {\em {Einstein Spaces}}.
\newblock Pergamon Press, 1969.

\bibitem{exac}
E.~Hertl, C.~Hoenselaers, D.~Kramer, M.~Maccallum, and H.~Stephani, {\em {Exact
  solutions of Einstein's field equations}}.
\newblock Cambridge Univ. Pr., 2003.

\bibitem{TheoLib:GEN15}
L.~Landau and E.~Lifshitz, {\em Fluid Mechanics}, vol.~6 of {\em Course of
  Theoretical Physics}.
\newblock Pergamon Press, first~ed., 1975.

\bibitem{reynolds}
O.~Reynolds, ``{On the dynamical theory of incompressible viscous fluids and
  the determination of the criterion},'' {\em Phil. Trans. Roy. Soc. London}
  {\bfseries 186} (1894) 123--164.

\end{thebibliography}\endgroup
\bibliographystyle{utphys}

\end{document}